\documentclass[aps,prd,twocolumns,eqsecnum,preprintnumbers,showpacs,amsmath,amssymb]
{revtex4}

\usepackage{graphicx}
\usepackage{dcolumn}
\usepackage{bm}

\begin{document}

\title{A TEMPORARY VIOLATION OF COLOR GAUGE INVARIANCE \\ AS A SOURCE
OF THE JAFFE-WITTEN MASS GAP IN QCD}

\author{V. Gogokhia}
\email[]{gogohia@rmki.kfki.hu}

\affiliation{HAS, CRIP, RMKI, Depart. Theor. Phys., Budapest 114,
P.O.B. 49, H-1525, Hungary}

\date{\today}
\begin{abstract}
We propose to realize a mass gap in QCD by not imposing the
transversality condition on the full gluon self-energy, while
preserving the color gauge invariance condition for the full gluon
propagator. This is justified by the nonlinear and nonperturbative
dynamics of QCD. None of physical observables/processes in
low-energy QCD will be directly affected by such a temporary
violation of color gauge invariance/symmetry. No
truncations/approximations and no special gauge choice are made
for the regularized skeleton loop integrals, contributing to the
full gluon self-energy, which enters the Schwinger-Dyson equation
for the full gluon propagator. In order to make the existence of a
mass gap perfectly clear the corresponding subtraction procedure
is introduced. All this allows one to establish the general
structure of the full gluon propagator and the corresponding gluon
Schwinger-Dyson equation in the presence of a mass gap. It is
mainly generated by the nonlinear interaction of massless gluon
modes. The physical meaning of the mass gap is to be responsible
for the large-scale (low-energy/momentum), i.e., nonperturbative
structure of the true QCD vacuum. In the presence of a mass gap
two different types of solutions for the full gluon propagator are
possible. The massive solution leads to an effective gluon mass,
which explicitly depends on the gauge-fixing parameter. This
solution becomes smooth at small gluon momentum in the Landau
gauge. The general iteration solution is always severely singular
at small gluon momentum, i.e., the gluons remain massless, and
this does not depend on the gauge choice. We also formulate a
general method how to restore the transversality of the gluon
propagator relevant for nonperturbative QCD.

\end{abstract}

\pacs{ 11.15.Tk, 12.38.Lg}

\keywords{}

\maketitle

\section{Introduction}

Quantum Chromodynamics (QCD) \cite{1,2} is widely accepted as a
realistic quantum field gauge theory of strong interactions not
only at the fundamental (microscopic) quark-gluon level but at the
hadronic (macroscopic) level as well. This means that in principle
it should describe the properties of experimentally observed
hadrons in terms of experimentally never seen quarks and gluons,
i.e., to describe the hadronic word from first principles -- an
ultimate goal of any fundamental theory. But this is a formidable
task because of the color confinement phenomenon, the dynamical
mechanism of which is not yet understood, and therefore the
confinement problem remains unsolved up to the present days. It
prevents colored quarks and gluons to be experimentally detected
as physical ("in" and "out" asymptotic) states which are colorless
(i.e., color-singlets), by definition, so color confinement is
permanent and absolute \cite{1}.

Today there is no doubt left that color confinement and other
dynamical effects, such as spontaneous breakdown of chiral
symmetry, bound-state problems, etc., being essentially
nonperturbative (NP) effects, are closely related to the
large-scale (low-energy/momentum) structure of the true QCD ground
state and vice-versa \cite{3,4} (and references therein). The
perturbation theory (PT) methods in general fail to investigate
them. If QCD itself is a confining theory then a characteristic
scale has to exist. It should be directly responsible for the
above-mentioned structure of the true QCD vacuum in the same way
as $\Lambda_{QCD}$ is responsible for the nontrivial perturbative
dynamics there (asymptotic freedom (AF) \cite{1}).

However, the Lagrangian of QCD \cite{1,2} does not contain
explicitly any of the mass scale parameters which could have a
physical meaning even after the corresponding renormalization
program is performed. So the main goal of this paper is to show
how the characteristic scale (the mass gap, for simplicity)
responsible for the NP dynamics in the infrared (IR) region may
explicitly appear in QCD. This becomes an imperative especially
after Jaffe and Witten have formulated their theorem "Yang-Mills
Existence And Mass Gap" \cite{5}. We will show that the mass gap
is dynamically generated mainly due to the nonlinear (NL)
interaction of massless gluon modes.

The propagation of gluons is one of the main dynamical effects in
the true QCD vacuum. It is described by the corresponding quantum
equation of motion, the so-called Schwinger-Dyson (SD) equation
\cite{1} (and references therein) for the full gluon propagator.
The importance of this equation is due to the fact that its
solutions reflect the quantum-dynamical structure of the true QCD
ground state. The color gauge structure of this equation is the
main subject of our investigation in order to find a way how to
realize a mass gap in QCD. Also we will discuss at least two
possible types of solutions of the gluon SD equation in the
presence of a mass gap, making no approximations/truncations and
no special gauge choice for the skeleton loop integrals
contributing to it. So they can be considered as the
generalizations of the explicit solutions because the latter ones
are necessarily based on the above-mentioned specific
approximations/truncations schemes.

\section{QED}

It is instructive to begin with a brief explanation why a mass gap
does not occur in quantum electrodynamics (QED). The photon SD
equation can be symbolically written down as follows:

\begin{equation}
D(q) = D^0(q) + D^0(q) \Pi(q) D(q),
\end{equation}
where we omit, for convenience, the dependence on the Dirac
indices, and $D^0(q)$ is the free photon propagator. $\Pi(q)$
describes the electron skeleton loop contribution to the photon
self-energy (the so-called vacuum polarization tensor).
Analytically it looks

\begin{equation}
\Pi(q) \equiv \Pi_{\mu\nu}(q) = - g^2 \int {i d^4 p \over (2
\pi)^4} Tr [\gamma_{\mu} S(p-q) \Gamma_{\nu}(p-q, q)S(p)],
\end{equation}
where $S(p)$ and $\Gamma_{\mu}(p-q,q)$ represent the full electron
propagator and the full electron-photon vertex, respectively. Here
and everywhere below the signature is Euclidean, since it implies
$q_i \rightarrow 0$ when $q^2 \rightarrow 0$ and vice-versa. This
tensor has the dimensions of a mass squared, and therefore it is
quadratically divergent. To make the formal existence of a mass
gap (the quadratically divergent constant, so having the
dimensions of a mass squared) perfectly clear, let us now, for
simplicity, subtract its value at zero. One obtains

\begin{equation}
\Pi^s(q) \equiv \Pi^s_{\mu\nu}(q) = \Pi_{\mu\nu}(q) -
\Pi_{\mu\nu}(0) = \Pi_{\mu\nu}(q) - \delta_{\mu\nu}\Delta^2
(\lambda).
\end{equation}
The explicit dependence on the dimensionless ultraviolet (UV)
regulating parameter $\lambda$ has been introduced into the mass
gap $\Delta^2(\lambda)$, given by the integral (2.2) at $q^2=0$,
in order to assign a mathematical meaning to it. In this
connection a few remarks are in order in advance.
 The dependence on $\lambda$ (when it is not shown
explicitly) is assumed in all divergent integrals here and below
in the case of the gluon self-energy as well (see next section).
This means that all the expressions are regularized (including
photon/gluon propagator), and we can operate with them as with
finite quantities. $\lambda$ should be removed on the final stage
only after performing the corresponding renormalization program
(which is beyond the scope of the present investigation, of
course). Whether the regulating parameter $\lambda$ has been
introduced in a gauge-invariant way (though this always can be
achieved) or not, and how it should be removed is not important
for the problem if a mass gap can be "released/liberated" from the
corresponding vacuum. We will show in the most general way (not
using the PT and no special gauge choice will be made) that this
is impossible in QED and might be possible in QCD.

The decomposition of the subtracted vacuum polarization tensor
into the independent tensor structures can be written as follows:

\begin{equation}
\Pi^s_{\mu\nu}(q) = T_{\mu\nu}(q) q^2 \Pi^s_1(q^2) + q_{\mu}
q_{\nu}(q) \Pi^s_2(q^2),
\end{equation}
where both invariant functions $\Pi^s_n(q^2)$ at $n=1,2$ are, by
definition, dimensionless and regular at small $q^2$, since
$\Pi^s(0) =0$; otherwise they remain arbitrary. From this relation
it follows that $\Pi^s(q) = O(q^2)$, i.e., it is always of the
order $q^2$. Also, here and everywhere below

\begin{equation}
T_{\mu\nu}(q)=\delta_{\mu\nu}-q_{\mu} q_{\nu} / q^2 =
\delta_{\mu\nu } - L_{\mu\nu}(q).
\end{equation}

Taking into account the subtraction (2.3), the photon SD equation
becomes

\begin{equation}
D(q) = D^0(q) + D^0(q) \Pi^s(q) D(q) + D^0(q) \Delta^2(\lambda)
D(q).
\end{equation}
Its subtracted part can be summed up into the geometric series, so
one has

\begin{equation}
D(q) = \tilde{D}^0(q) + \tilde{D}^0(q) \Delta^2(\lambda) D(q),
\end{equation}
where the modified photon propagator is

\begin{equation}
\tilde{D}^0(q) = {D^0(q) \over  1 - \Pi^s(q) D^0(q)}= D^0(q) +
D^0(q) \Pi^s(q) D^0(q) - D^0(q)\Pi^s(q)D^0(q) \Pi^s(q)D^0(q) + ...
\ .
\end{equation}
Since $\Pi^s(q) = O(q^2)$ and $D^0(q) \sim (q^2)^{-1}$, the IR
singularity of the modified photon propagator is determined by the
IR singularity of the free photon propagator, i.e.,
$\tilde{D}^0(q) = O (D^0(q))$ with respect to the behavior at
small photon momentum.

Similar to the subtracted photon self-energy, the photon
self-energy (2.2) in terms of independent tensor structures is

\begin{equation}
\Pi_{\mu\nu}(q) = T_{\mu\nu}(q) q^2 \Pi_1(q^2) + q_{\mu} q_{\nu}
\Pi_2(q^2),
\end{equation}
where again $\Pi_n(q^2)$ at $n=1,2$ are dimensionless functions
and remain arbitrary. Due to the transversality of the photon
self-energy

\begin{equation}
q_{\mu} \Pi_{\mu\nu}(q) =q_{\nu} \Pi_{\mu\nu}(q) =0,
\end{equation}
which comes from the current conservation condition in QED, one
has $\Pi_2(q^2)=0$, i.e., it should be purely transversal

\begin{equation}
\Pi_{\mu\nu}(q) = T_{\mu\nu}(q) q^2 \Pi_1(q^2).
\end{equation}
On the other hand, from the subtraction (2.3) and the
transversality condition (2.10) it follows that

\begin{equation}
\Pi^s_2(q^2)= - {\Delta^2(\lambda) \over q^2},
\end{equation}
which, however, is impossible since $\Pi^s_2(q^2)$ is a regular
function of $q^2$, by definition. So the mass gap should be zero
and consequently $\Pi^s_2(q^2)$ as well, i.e.,

\begin{equation}
\Pi^s_2(q^2)=0, \quad \Delta^2(\lambda)=0.
\end{equation}
Thus the subtracted photon self-energy is also transversal, i.e.,
satisfies the transversality condition

\begin{equation}
q_{\mu} \Pi_{\mu\nu}(q) =q_{\nu} \Pi^s_{\mu\nu}(q) =0,
\end{equation}
and coincides with the photon self-energy (see Eq. (2.3) at the
zero mass gap). Moreover, this means that the photon self-energy
does not have a pole in its invariant function $\Pi_1(q^2)=
\Pi^s_1(q^2)$. As mentioned above, in obtaining these results
neither the PT has been used nor a special gauge has been chosen.
So there is no place for quadratically divergent constants in QED,
while logarithmic divergence still can be present in the invariant
function $\Pi_1(q^2) = \Pi^s_1(q^2)$. It is to be included into
the electric charge through the corresponding renormalization
program (for these detailed gauge-invariant derivations explicitly
done in lower order of the PT see Refs. \cite{2,6,7,8,9}).

In fact, the current conservation condition (2.10) lowers the
quadratical divergence of the corresponding integral (2.2) to a
logarithmic one. That is the reason why in QED logarithmic
divergences survive only. Thus in QED there is no mass gap and the
relevant photon SD equation is shown in Eq. (2.8), simply
identifying the full photon propagator with its modified
counterpart. In other words, in QED we can replace $\Pi(q)$ by its
subtracted counterpart $\Pi^s(q)$ from the very beginning ($\Pi(q)
\rightarrow \Pi^s(q)$), totally discarding the quadratically
divergent constant $\Delta^2(\lambda)$ from all the equations and
relations. The current conservation condition for the photon
self-energy (2.10), i.e., its transversality, and for the full
photon propagator $q_{\mu}q_{\nu}D_{\mu\nu}(q) = i\xi$, where
$\xi$ is the gauge-fixing parameter, are consequences of gauge
invariance. They should be maintained at every stage of the
calculations, since the photon is a physical state. In other
words, at all stages the current conservation plays a crucial role
in extracting physical information from the $S$-matrix elements in
QED. For example, if some QED process includes the full photon
propagator, then the corresponding $S$-matrix element is
proportional to the combination $j^{\mu}_1 (q)D_{\mu\nu}(q)
j^{\nu}_2(q)$. The current conservation condition $j^{\mu}_1 (q)
q_{\mu} = j^{\nu}_2(q)q_{\nu} =0$ implies that the unphysical
(longitudinal) component of the full photon propagator does not
change the physics of QED, i.e., only its physical (transversal)
component is important. In its turn this means that the
transversality condition imposed on the photon self-energy is
important, because $\Pi_{\mu\nu}(q)$ itself is a correction to the
amplitude of the physical process, for example such as
electron-electron scattering.

\section{QCD}

Due do color confinement in QCD the gluon is not a physical state.
Still, color gauge invariance should also be preserved, so the
color current conservation takes place in QCD as well. However, in
this theory it plays no role in the extraction of physical
information from the $S$-matrix elements for the corresponding
physical processes and quantities. So in QCD there is no such
physical amplitude to which the gluon self-energy may directly
contribute (for example, quark-quark/antiquark scattering is not a
physical process). The lesson which comes from QED is that if one
preserves the transversality of the photon self-energy at every
stage, then there is no mass gap. Thus, in order to realize a mass
gap in QCD, our proposal is not to impose the transversality
condition on the gluon self-energy, but preserving the color gauge
invariance condition for the full gluon propagator (see below). As
mentioned above, no QCD physics will be directly affected by this.
So color gauge symmetry will be violated at the initial stage (at
the level of the gluon self-energy) and will be restored at the
final stage (at the level of the full gluon propagator).

\subsection{Gluon SD equation}

The gluon SD equation can be symbolically written down as follows
(for our purposes it is more convenient to consider the SD
equation for the full gluon propagator and not for its inverse):

\begin{equation}
D_{\mu\nu}(q) = D^0_{\mu\nu}(q) + D^0_{\mu\rho}(q) i
\Pi_{\rho\sigma}(q; D) D_{\sigma\nu}(q),
\end{equation}
where $D^0_{\mu\nu}(q)$ is the free gluon propagator.
$\Pi_{\rho\sigma}(q; D)$ is the gluon self-energy, and in general
it depends on the full gluon propagator due to the non-Abelian
character of QCD (see below). Thus the gluon SD equation is highly
NL, while the photon SD equation (2.1) is a linear one. In what
follows we omit the color group indices, since for the gluon
propagator (and hence for its self-energy) they are reduced to the
trivial $\delta$-function, for example $D^{ab}_{\mu\nu}(q) =
D_{\mu\nu}(q)\delta^{ab}$. Also, for convenience, we introduce $i$
into the gluon SD equation (3.1).

The gluon self-energy $\Pi_{\rho\sigma}(q; D)$ is the sum of a few
terms, namely

\begin{equation}
 \Pi_{\rho\sigma}(q; D)= - \Pi^q_{\rho\sigma}(q) -
\Pi^{gh}_{\rho\sigma}(q) + \Pi_{\rho\sigma}^t(D) +
\Pi_{(1)\rho\sigma}(q; D) + \Pi_{(2)\rho\sigma}(q; D) +
\Pi'_{(2)\rho\sigma}(q; D),
\end{equation}
where $\Pi^q_{\rho\sigma}(q)$ describes the skeleton loop
contribution due to quark degrees of freedom (it is an analog of
the vacuum polarization tensor in QED, see Eq. (2.2)), while
$\Pi^{gh}_{\rho\sigma}(q)$ describes the skeleton loop
contribution due to ghost degrees of freedom. Both skeleton loop
integrals do not depend on the full gluon propagator $D$, so they
represent the linear contribution to the gluon self-energy.
$\Pi_{\rho\sigma}^t(D)$ represents the so-called constant skeleton
tadpole term. $\Pi_{(1)\rho\sigma}(q; D)$ represents the skeleton
loop contribution, which contains the triple gluon vertices only.
$\Pi_{(2)\rho\sigma}(q; D)$ and $\Pi'_{(2)\rho\sigma}(q; D)$
describe topologically independent skeleton two-loop
contributions, which combine the triple and quartic gluon
vertices. The last four terms explicitly contain the full gluon
propagators in different powers, that is why they form the NL part
of the gluon self-energy. The explicit expressions for the
corresponding skeleton loop integrals \cite{10} (in which the
corresponding symmetry coefficients can be included) are of no
importance here. Let us note that like in QED these skeleton loop
integrals are in general quadratically divergent, and therefore
they should be regularized (see remarks above and below).

\subsection{A temporary violation of color gauge
invariance/symmetry (TVCGI/S)}

The color gauge invariance condition for the gluon self-energy
(3.2) can be reduced to the three independent transversality
conditions imposed on it. It is well known that the quark
contribution can be made transversal independently of the pure
gluon contributions within any regularization scheme which
preserves gauge invariance, for example such as the dimensional
regularization method (DRM) \cite{1,2,8,9,11}. So, one has

\begin{equation}
q_{\rho} \Pi^q_{\rho\sigma}(q) =q_{\sigma} \Pi^q_{\rho\sigma}(q) =
0,
\end{equation}
indeed. In the same way the sum of the gluon contributions can be
done transversal by taking into account the ghost contribution, so
again one has

\begin{equation}
q_{\rho} \Bigl[ \Pi_{(1)\rho\sigma}(q; D) + \Pi_{(2)\rho\sigma}(q;
D) + \Pi'_{(2)\rho\sigma}(q; D) - \Pi^{gh}_{\rho\sigma}(q) \Bigr]
= 0.
\end{equation}
The role of ghost degrees of freedom is to cancel the unphysical
(longitudinal) component of gauge bosons in every order of the PT,
i.e., going beyond the PT and thus being general. The previous
equation just demonstrates this, since it contains the
corresponding skeleton loop integrals.

However, there is no such regularization scheme (preserving or not
gauge invariance) in which the transversality condition for the
constant skeleton tadpole term could be satisfied, i.e.,

\begin{equation}
q_{\rho} \Pi^t_{\rho\sigma}(D) = q_{\rho} \delta_{\rho\sigma}
\Delta^2_t(D) = q_{\sigma} \Delta^2_t(D) \neq 0,
\end{equation}
indeed. This means that in any NP approach the transversality
condition imposed on the gluon self-energy may not be valid, i.e.,
in general

\begin{equation}
q_{\rho} \Pi_{\rho\sigma}(q; D) =q_{\sigma} \Pi_{\rho\sigma}(q; D)
\neq 0.
\end{equation}
In the PT, when the full gluon propagator is always approximated
by the free one, the constant tadpole term is set to be zero
within the DRM \cite{8,11}, i.e., $\Pi^t_{\rho\sigma}(D^0) =0$. So
in the PT the transversality condition for the gluon self-energy
is always satisfied.

The relation (3.6) justifies our proposal not to impose the
transversality condition on the gluon self-energy. The relation
(3.5) emphasizes the special role of the constant skeleton tadpole
term in the NP QCD dynamics. It explicitly violates the
transversality condition for the gluon self-energy (3.6). The
second important observation is that now ghosts themselves cannot
automatically provide the transversality of the gluon propagator
in NP QCD. However, this does not mean that we need no ghosts at
all. Of course, we need them in other sectors of QCD, for example
in the quark-gluon Ward-Takahashi identity, which contains the
so-called ghost-quark scattering kernel explicitly \cite{1}.

\subsection{Subtractions}

As we already know from QED, the regularization of the gluon
self-energy can be started from the subtraction its value at the
zero point (see, however, remarks below). Thus, quite similarly to
the subtraction (2.3), one obtains

\begin{equation}
\Pi^s_{\rho\sigma}(q; D) = \Pi_{\rho\sigma}(q; D) -
\Pi_{\rho\sigma}(0; D) = \Pi_{\rho\sigma}(q; D) -
\delta_{\rho\sigma} \Delta^2 (\lambda; D).
\end{equation}
Let us remind once more that for our purpose, namely to
demonstrate a possible existence of a mass gap $\Delta^2 (\lambda;
D)$ in QCD, it is not important how $\lambda$ has been introduced
and  how it should be removed at the final stage. The mass gap
itself is mainly generated by the nonlinear interaction of
massless gluon modes, slightly corrected by the linear
contributions coming from the quark and ghost degrees of freedom,
namely

\begin{equation}
\Delta^2 (\lambda; D)= \Pi^t(D) + \sum_a \Pi^a(0; D) =
\Delta^2_t(D) + \sum_a \Delta^2_a(0; D) ,
\end{equation}
where index "a" runs as follows: $a= -q, -gh, 1, 2, 2'$, and $-q,
\ - gh$ mean that both terms enter the above-mentioned sum with
minus sign (here, obviously, the tensor indices are omitted). In
these relation all the divergent constants $\Pi^t(D)$ and
$\Pi^a(0; D)$, having the dimensions of a mass squared, are given
by the corresponding skeleton loop integrals at $q^2=0$. Thus
these constants summed up into the mass gap squared (3.8) cannot
be discarded like in QED, since the transversality condition for
the gluon self-energy is not satisfied, see Eq. (3.6). In other
words, in QCD in general the quadratical divergences of the
corresponding loop integrals cannot be lowered to logarithmic
ones, and therefore the mass gap (3.8) should be explicitly taken
into account in this theory. The transversality condition for the
gluon self-energy can be satisfied partially, i.e., if one imposes
it on quark and gluon (along with ghost) degrees of freedom as it
follows from above. Then the mass gap is to be reduced to
$\Pi^t(D)$, since all other constants $\Pi^a(0;D)$ can be
discarded in this case (see Eq. (3.8)). However, we will stick to
our proposal not to impose the transversality condition on the
gluon self-energy, and thus to deal with the mass gap on account
of all possible contributions.

The subtracted gluon self-energy

\begin{equation}
\Pi^s_{\rho\sigma}(q; D) \equiv \Pi^s(q; D) = \sum_a \Pi^s_a(q; D)
\end{equation}
is free from the tadpole contribution, because $\Pi^s_t(D) =
\Pi_t(D)- \Pi_t(D)=0$, by definition, at any $D$, while in the
gluon self-energy it is explicitly present through the mass gap
(see Eqs. (3.8) and (3.7)). The general decomposition of the
subtracted gluon self-energy into the independent tensor
structures can be written down as follows:

\begin{equation}
\Pi^s_{\rho\sigma}(q; D) = T_{\rho\sigma}(q) q^2 \Pi(q^2; D) +
q_{\rho} q_{\sigma} \tilde{\Pi}(q^2; D),
\end{equation}
where both invariant functions $\Pi(q^2; D)$ and $\tilde{\Pi}(q^2;
D)$ are dimensionless and regular at small $q^2$. Since the
subtracted gluon self-energy does not contain the tadpole
contribution, we can now impose the color current conservation
condition on it, i.e., to put

\begin{equation}
q_{\rho} \Pi^s_{\rho\sigma}(q; D)= q_{\sigma}
\Pi^s_{\rho\sigma}(q; D) = 0,
\end{equation}
which implies $\tilde{\Pi}(q^2; D) = 0$, so that the subtracted
gluon self-energy finally becomes purely transversal

\begin{equation}
\Pi^s_{\rho\sigma}(q; D) = T_{\rho\sigma}(q) q^2 \Pi(q^2; D),
\end{equation}
and it is always of the order $q^2$ at any $D$, since the
invariant function $\Pi(q^2; D)$ is regular at small $q^2$ at any
$D$. Thus the subtracted quantities are free from the quadratic
divergences, but logarithmic ones can be still present in
$\Pi(q^2; D)$ like in QED.

\subsection{General structure of the gluon SD equation}

Our strategy is not to impose the transversality condition on the
gluon self-energy in order to realize a mass gap despite whether
or not the tadpole term is explicitly present. To show that this
works, it is instructive to substitute the subtracted gluon
self-energy (3.10) (and not its transversal part (3.12)) into the
initial gluon SD equation (3.1), on account of the subtraction
(3.7). Then one obtains

\begin{equation}
D_{\mu\nu}(q) = D^0_{\mu\nu}(q) + D^0_{\mu\rho}(q)i[
T_{\rho\sigma}(q) q^2 \Pi(q^2; D) + q_{\rho}q_{\sigma}
\tilde{\Pi}(q^2; D)]D_{\sigma\nu}(q) + D^0_{\mu\sigma}(q)i
\Delta^2(\lambda; D) D_{\sigma\nu}(q).
\end{equation}
Let us now introduce the general tensor decompositions of the full
and auxiliary free gluon propagators
$D_{\mu\nu}(q)=i[T_{\mu\nu}(q) d(q^2) +
L_{\mu\nu}(q)d_1(q^2)](1/q^2)$ and

\begin{equation}
D^0_{\mu\nu}(q)=i[ T_{\mu\nu}(q) + L_{\mu\nu}(q) d_0(q^2)](1/q^2),
\end{equation}
respectively. The form factor $d_0(q^2)$ introduced into the
unphysical part of the auxiliary free gluon propagator
$D^0_{\mu\nu}(q)$ is needed in order to explicitly show that the
longitudinal part of the subtracted gluon self-energy
$\tilde{\Pi}(q^2; D)$ plays no role. The color gauge invariance
condition imposed on the full gluon propagator

\begin{equation}
q_{\mu}q_{\nu}D_{\mu\nu}(q) = i \xi,
\end{equation}
implies $d_1(q^2) = \xi$, so that the full gluon propagator
becomes

\begin{equation}
D_{\mu\nu}(q) = i \left\{ T_{\mu\nu}(q) d(q^2) + \xi L_{\mu\nu}(q)
\right\} {1 \over q^2}.
\end{equation}
Substituting all these decompositions into the gluon SD equation
(3.13), one obtains

\begin{equation}
d(q^2) = {1 \over 1 + \Pi(q^2; D) + (\Delta^2(\lambda; D) / q^2)},
\end{equation}
and

\begin{equation}
d_0(q^2) = {\xi \over 1 - \xi [\tilde{\Pi}(q^2; D) +
(\Delta^2(\lambda; D) / q^2)]}.
\end{equation}
However, the auxiliary free gluon propagator defined in Eqs.
(3.14) and (3.18) is to be equivalently replaced as follows:

\begin{equation}
D^0_{\mu\nu}(q) \Longrightarrow D^0_{\mu\nu}(q) + i \xi
L_{\mu\nu}(q) d_0(q^2) \Bigl[ \tilde{\Pi}(q^2; D) + {
\Delta^2(\lambda; D) \over q^2} \Bigr] {1 \over q^2},
\end{equation}
where $D^0_{\mu\nu}(q)$ in the right-hand-side is the standard
free gluon propagator, i.e.,

\begin{equation}
D^0_{\mu\nu}(q) = i \left\{ T_{\mu\nu}(q) + \xi L_{\mu\nu}(q)
\right\} {1 \over q^2}.
\end{equation}
Then the gluon SD equation in the presence of the mass gap (3.13),
on account of the explicit expression for the auxiliary free gluon
form factor (3.18), and doing some tedious algebra, is also to be
equivalently replaced as follows:

\begin{eqnarray}
D_{\mu\nu}(q) &=& D^0_{\mu\nu}(q) + D^0_{\mu\rho}(q)i
T_{\rho\sigma}(q) q^2 \Pi(q^2; D) D_{\sigma\nu}(q) \nonumber\\
&+& D^0_{\mu\sigma}(q)i \Delta^2(\lambda; D) D_{\sigma\nu}(q) + i
\xi^2 L_{\mu\nu}(q) { \Delta^2(\lambda; D) \over q^4}.
\end{eqnarray}
Here and below $D^0_{\mu\nu}(q)$ is the free gluon
propagator(3.20). The gluon SD equation (3.21) does not depend on
$d_0(q^2)$ and $\tilde{\Pi}(q^2; D)$, i.e., they played their role
and then retired from the scene. So, our derivation explicitly
shows that the longitudinal part of the subtracted gluon
self-energy $\tilde{\Pi}(q^2; D)$ plays no role and can be put to
zero without loosing generality, and thus making the subtracted
gluon self-energy purely transversal in accordance with Eq.
(3.12).

Using now the explicit expression for the free gluon propagator
(3.20) this equation can be further simplified to

\begin{equation}
D_{\mu\nu}(q) = D^0_{\mu\nu}(q) - T_{\mu\sigma}(q) \Bigl[\Pi(q^2;
D) + { \Delta^2(\lambda; D) \over q^2} \Bigr] D_{\sigma\nu}(q).
\end{equation}
It is easy to check that the full gluon propagator satisfies the
color gauge invariance condition (3.15), indeed. So the full gluon
propagator is the expression (3.16) with the full gluon form
factor given in Eq. (3.17), which obviously satisfies Eq. (3.22).
The only price we have paid by violating color gauge invariance is
the gluon self-energy, while the full and free gluon propagators
and the subtracted gluon self-energy always satisfy it. Let us
emphasize that the expression for the full gluon form factor shown
in the relation (3.17) cannot be considered as the formal solution
for the full gluon propagator, since both the mass gap
$\Delta^2(\lambda; D)$ and the invariant function $\Pi(q^2; D)$
depend on $D$ themselves. Here it is worth noting in advance that
from above it is almost clear that if one begins with the UV
renormalization program, then the information on the mass gap will
be totally lost. In this case instead of the regularized gluon
self-energy its subtracted regularized counterpart comes into the
play. In other words, in the PT limit $\Delta^2(\lambda; D)=0$ one
recovers the standard gluon SD equation, and the gluon self-energy
coincides with its subtracted counterpart like in QED. For a more
detailed explanation see below subsection C in section V.

Thus, we have established the general structure of the full gluon
propagator (see Eqs. (3.16) and (3.17)) and the corresponding
gluon SD equation (3.22) (which is equivalent to Eq. (3.13)) in
the presence of a mass gap.

\section{Massive solution}

An immediate consequence of the explicit presence of the mass gap
in the full gluon propagator is that a massive-type solution for
it becomes possible. In other words, in this case the gluon may
indeed acquire an effective mass. From Eq. (3.17) it follows that

\begin{equation}
{ 1 \over q^2} d(q^2) = {1 \over q^2 + q^2 \Pi(q^2; \xi) +
\Delta^2(\lambda, \xi)},
\end{equation}
where instead of the dependence on $D$ the dependence on $\xi$ is
explicitly shown. The full gluon propagator (3.16) may have a
pole-type solution at the finite point if and only if the
denominator in Eq. (4.1) has a zero at this point $q^2 = - m^2_g$
(Euclidean signature), i.e.,

\begin{equation}
- m^2_g  - m^2_g \Pi(-m^2_g; \xi) + \Delta^2(\lambda, \xi)=0,
\end{equation}
where $m^2_g \equiv m^2_g(\lambda, \xi)$ is an effective gluon
mass, and the previous equation is a transcendental equation for
its determination. Excluding the mass gap, one obtains that the
denominator in the full gluon propagator becomes

\begin{equation}
q^2 + q^2 \Pi(q^2; \xi) + \Delta^2(\lambda, \xi) = q^2 + m^2_g +
q^2 \Pi(q^2; \xi) + m^2_g \Pi(-m^2_g; \xi).
\end{equation}

Let us now expand $\Pi(q^2; \xi)$ in a Taylor series near $m^2_g$:

\begin{equation}
\Pi(q^2; \xi) = \Pi(-m^2_g; \xi) + (q^2 + m^2_g) \Pi'(-m^2_g; \xi)
+ O \Bigl( (q^2 + m^2_g)^2 \Bigr).
\end{equation}
Substituting this expansion into the previous relation and after
doing some tedious algebra, one obtains

\begin{equation}
q^2 + m^2_g + q^2 \Pi(q^2; \xi) + m^2_g \Pi(-m^2_g; \xi)= (q^2 +
m^2_g)[1 +  \Pi(-m^2_g; \xi) - m^2_g \Pi'(-m^2_g; \xi)] [ 1 +
\Pi^R(q^2; \xi)],
\end{equation}
where $\Pi^R(q^2; \xi)= 0$ at $q^2=-m^2_g$; otherwise it remains
arbitrary. Thus the full gluon propagator (3.16) now looks

\begin{equation}
D_{\mu\nu}(q) = i T_{\mu\nu}(q) {Z_3 \over (q^2 + m^2_g) [ 1 +
\Pi^R(q^2; m^2_g)]} + i \xi L_{\mu\nu}(q) {1 \over q^2},
\end{equation}
where, for future purpose, in the invariant function $\Pi^R(q^2;
m^2_g)$ instead of $\xi$ we introduced the dependence on the gluon
effective mass squared $m_g^2$ which depends on $\xi$ itself. The
gluon renormalization constant is

\begin{equation}
Z_3 = [1 + \Pi(-m^2_g; \xi) - m^2_g \Pi'(-m^2_g; \xi)]^{-1}.
\end{equation}
In the formal PT limit $\Delta^2(\lambda, \xi) =0$, an effective
gluon mass is also zero, $m_g^2(\lambda, \xi) =0$, as it follows
from Eq. (4.2). So an effective gluon mass is the NP effect. At
the same time, it cannot be interpreted as the "physical" gluon
mass, since it remains explicitly gauge-dependent quantity. The
gluon renormalization constant (4.7) in this limit becomes a
standard one, namely $[1 + \Pi(0; \xi)]^{-1}$. The massive-type
solution (4.6) becomes smooth in the IR ($q^2 \rightarrow 0$) in
the Landau gauge $\xi=0$ only (the ghosts now cannot guarantee the
cancellation of the longitudinal part of the full gluon propagator
as mentioned above). In this connection let us point out that
Landau gauge smooth (even vanishing in the IR) gluon propagator at
the expense of more singular (than the free one) in the IR ghost
propagator has been obtained and discussed in Refs. \cite{12,13}
(and references therein). As mentioned above, however, these
results are necessarily based on different
approximations/truncations for the skeleton loop integrals
contributing to the gluon self-energy.

It is interesting to note that Eq. (4.2) has a second solution in
the PT limit $\Delta^2(\lambda, \xi) =0$. In this case, an
effective gluon mass remains finite, but $1 + \Pi(-m^2_g; \xi)
=0$. So a scale responsible for the NP dynamics is not determined
by the gluon mass itself, but by the condition $1 + \Pi(-m^2_g;
\xi) =0$. Its interpretation from a physical point of view is not
clear.

\section{Iteration solution}

The expression for the full gluon form factor shown in the
relation (3.17) cannot be considered as the formal solution for
the full gluon propagator, since both the mass gap
$\Delta^2(\lambda; D)$ and the invariant function $\Pi(q^2; D)$
depend on $D$ themselves. In order to perform a formal iteration
of the gluon SD equation (3.22), much more convenient to address
to its "solution" for the full gluon form factor (3.17),
nevertheless, and rewrite it as follows:

\begin{equation}
d(q^2) = 1 - \Bigl[ \Pi(q^2; d) + {\Delta^2(\lambda; d) \over q^2}
\Bigr] d(q^2) = 1 - P(q^2; d) d(q^2),
\end{equation}
i.e., in the form of the corresponding transcendental (i.e., not
algebraic) equation suitable for the formal nonlinear iteration
procedure. Here we replace the dependence on $D$ by the equivalent
dependence on $d$. For future purposes, it is convenient to
introduce short-hand notations as follows:

\begin{eqnarray}
\Delta^2(\lambda; d=d^{(0)} + d^{(1)} + d^{(2)} + ... + d^{(m)}+
...
) &=& \Delta^2_m = \Delta^2 c_m(\lambda, \alpha, \xi, g^2), \nonumber\\
\Pi(q^2; d=d^{(0)} + d^{(1)}+d^{(2)}+ ... + d^{(m)} + ...) &=&
\Pi_m(q^2),
\end{eqnarray}
and

\begin{equation}
P_m(q^2) = \Bigl[ \Pi_m(q^2) + {\Delta^2_m \over q^2} \Bigr], \
m=0,1,2,3,... \ .
\end{equation}
In these relations $\Delta^2_m$ are the auxiliary mass squared
parameters, while $\Delta^2$ is the mass gap itself (see, however,
remarks in Conclusions). The dimensionless constants $c_m$ via the
corresponding subscripts depend on which iteration for the gluon
form factor $d$ is actually done. They may depend on the
dimensionless coupling constant squared $g^2$, as well as on the
gauge-fixing parameter $\xi$. We also introduce the explicit
dependence on the dimensionless finite (slightly different from
zero) subtraction point $\alpha$, since the initial subtraction at
the zero point may be dangerous \cite{1}. The dependence of
$\Delta^2$ on all these parameters is not shown explicitly, and if
necessary can be restored any time. Let us also remind that all
the invariant functions $\Pi_m(q^2)$ are regular at small $q^2$.
If it were possible to express the full gluon form factor $d(q^2)$
in terms of these quantities then it would be the formal solution
for the full gluon propagator. In fact, this is nothing but the
skeleton loops expansion, since the regularized skeleton loop
integrals, contributing to the gluon self-energy, have to be
iterated. This is the so-called general iteration solution. No
truncations/approximations and no special gauge choice have been
made. This formal expansion is not a PT series. The magnitude of
the coupling constant squared and the dependence of the
regularized skeleton loop integrals on it is completely arbitrary.

It is instructive to describe the general iteration procedure in
some details. Evidently, $d^{(0)}=1$, and this corresponds to the
approximation of the full gluon propagator by its free counterpart
in the gluon SD equation (3.22). Doing the first iteration in Eq.
(5.1), one thus obtains

\begin{equation}
d(q^2) = 1 - P_0(q^2) + ... = 1 + d^{(1)}(q^2) + ...,
\end{equation}
where obviously

\begin{equation}
d^{(1)}(q^2) = - P_0(q^2).
\end{equation}
Doing the second iteration, one obtains

\begin{equation}
d(q^2) = 1 - P_1(q^2) [ 1 + d^{(1)}(q^2) ] + ... = 1 +
d^{(1)}(q^2) + d^{(2)}(q^2) + ...,
\end{equation}
where

\begin{equation}
d^{(2)}(q^2) = - d^{(1)}(q^2) - P_1(q^2) [ 1 - P_0(q^2)].
\end{equation}
Doing the third iteration, one further obtains

\begin{equation}
d(q^2) = 1 - P_2(q^2) [ 1 + d^{(1)}(q^2) + d^{(2)}(q^2)] + ... = 1
+ d^{(1)}(q^2) + d^{(2)}(q^2) + d^{(3)}(q^2) + ...,
\end{equation}
where

\begin{equation}
d^{(3)}(q^2) = - d^{(1)}(q^2) - d^{(2)}(q^2) - P_2(q^2) [ 1 -
P_1(q^2)(1 - P_0(q^2))],
\end{equation}
and so on for the next iterations.

Thus up to the third iteration, one finally obtains

\begin{equation}
d(q^2) = \sum_{m=0}^{\infty} d^{(m)}(q^2) = 1 - [\Pi_2(q^2) +
{\Delta^2_2 \over q^2}] \Bigl[ 1 - [\Pi_1(q^2) + {\Delta^2_1 \over
q^2}] [1 - \Pi_0(q^2) - {\Delta^2_0 \over q^2}] \Bigr] + ... \ .
\end{equation}
We restrict ourselves by the iterated gluon form factor up to the
third term, since this already allows to show explicitly some
general features of such kind of the nonlinear iteration
procedure.

\subsection{Splitting/shifting procedure}

Doing some tedious algebra, the previous expression can be
rewritten as follows:

\begin{eqnarray}
d(q^2) &=& [1 - \Pi_2(q^2) + \Pi_1(q^2) \Pi_2(q^2) - \Pi_0(q^2)
\Pi_1(q^2)\Pi_2(q^2) + ...] \nonumber\\
&+& {1 \over q^2} [\Pi_2(q^2)\Delta^2_1 + \Pi_1(q^2)\Delta^2_2 -
\Pi_0(q^2) \Pi_1(q^2)\Delta^2_2 - \Pi_0(q^2) \Pi_2(q^2)\Delta^2_1
- \Pi_1(q^2) \Pi_2(q^2)\Delta^2_2 + ...] \nonumber\\
&-& {1 \over q^4} [\Pi_0(q^2) \Delta^2_1 \Delta^2_2 + \Pi_1(q^2)
\Delta^2_0 \Delta^2_2 + \Pi_2(q^2) \Delta^2_0 \Delta^2_1 + ...]
\nonumber\\
&-& {1 \over q^2} [\Delta^2_2 -  {\Delta^2_1 \Delta^2_2 \over q^2}
+ { \Delta^2_0 \Delta^2_1 \Delta^2_2 \over q^4} + ...],
\end{eqnarray}
so that  this formal expansion contains three different types of
terms. The first type are the terms which contain only different
combinations of $\Pi_m(q^2)$ (they are not multiplied by inverse
powers of $q^2$); the third type of terms contains only different
combinations of $(\Delta^2_m / q^2)$. The second type of terms
contains the so-called mixed terms, containing the first and third
types of terms in different combinations. The two last types of
terms are multiplied by the corresponding powers of $1/q^2$.
Evidently, such structure of terms will be present in each
iteration term for the full gluon form factor. However, any of the
mixed terms can be split exactly into the first and third types of
terms by keeping the necessary number of terms in the Taylor
expansions in powers of $q^2$ for $\Pi_m(q^2)$, which are regular
functions at small $q^2$. Thus the IR structure of the full gluon
form factor (which just is our primary goal to establish) is
determined not only by the third type of terms. It gains
contributions from the mixed terms as well.

Let us present the above-mentioned Taylor expansions as follows:

\begin{equation}
\Pi_m(q^2) = \Pi_m(0) + (q^2 / \mu^2) \Pi^{(1)}_m (0) + (q^2 /
\mu^2)^2 \Pi^{(2)}_m (0) + O_m(q^6),
\end{equation}
since for the third iteration we need to use the Taylor expansions
up to this order (here $\mu^2$ is some fixed mass squared (not to
be mixed up with the tensor index)). For example, the mixed term
$(1/ q^2) \Pi_2(q^2)\Delta^2_1$ should be split as

\begin{equation}
{\Delta^2_1 \over q^2}\Pi_2(q^2) = {\Delta^2_1 \over q^2} \Bigl[
\Pi_2(0) + (q^2 / \mu^2) \Pi^{(1)}_2 (0) + O(q^4) \Bigr] =
{\Delta^2_1 \over q^2} \Pi_2(0) + a_1\Pi^{(1)}_2 (0) +O(q^2).
\end{equation}
Here and everywhere below $a_m = (\Delta^2_m / \mu^2), \
m=0,1,2,3,...$ are the dimensionless constants. The first term now
is to be shifted to the third type of terms and combined with the
term $(-1/q^2)\Delta^2_2$, while the second term $a_1\Pi^{(1)}_2
(0) +O(q^2)$ is to be shifted to the first type of terms. All
other mixed terms of similar structure should be treated
absolutely in the same way. For the mixed term $(-1 / q^4)
\Pi_0(q^2) \Delta^2_1 \Delta^2_2$, one has

\begin{eqnarray}
- {\Delta^2_1 \Delta^2_2 \over q^4}\Pi_0(q^2) &=& -{\Delta^2_1
\Delta^2_2 \over q^4} \Bigl[ \Pi_0(0) + (q^2 / \mu^2) \Pi^{(1)}_0
(0) +  (q^2 / \mu^2)^2 \Pi^{(2)}_0 (0) + O(q^6) \Bigr]
\nonumber\\
&=& - {\Delta^2_1 \Delta^2_2 \over q^4} \Pi_0(0) - {\Delta^2_1
\over q^2} a_2 \Pi^{(1)}_0 (0) - a_1a_2\Pi^{(2)}_0 (0)- O(q^2).
\end{eqnarray}
Again the first and second terms should be shifted to the third
type of terms and combined with terms containing there the same
powers of $1/q^2$, while the last two terms should be shifted to
the first type of terms.

Similar to the Taylor expansion (5.12), one has

\begin{equation}
\Pi_m(q^2)\Pi_n (q^2)= \Pi_{mn} (q^2) = \Pi_{mn}(0) + (q^2 /
\mu^2) \Pi^{(1)}_{mn} (0) + (q^2 / \mu^2)^2 \Pi^{(2)}_{mn} (0) +
O_{mn}(q^6).
\end{equation}
Then, for example the mixed term $(-1/q^2) \Pi_0(q^2)
\Pi_1(q^2)\Delta^2_2$ can be split as

\begin{eqnarray}
-{ \Delta^2_2 \over q^2} \Pi_0(q^2) \Pi_1(q^2) &=& -{ \Delta^2_2
\over q^2} \Bigl[ \Pi_{01}(0) + (q^2 / \mu^2) \Pi^{(1)}_{01} (0) +
O(q^4) \Bigr] \nonumber\\
&=& -{ \Delta^2_2 \over q^2} \Pi_{01}(0) - a_2 \Pi^{(1)}_{01} (0)
+ O(q^2),
\end{eqnarray}
so again the first term should be shifted to the third type of
terms and combined with the terms containing the corresponding
powers of $1/q^2$, while other terms are to be shifted to the
first type of terms.

Completing this exact splitting/shifting procedure in the
expansion (5.11), one can in general represent it as follows:

\begin{equation}
d(q^2) = \Bigl( {\Delta^2 \over q^2} \Bigr) B_1(\lambda, \alpha,
\xi, g^2) + \Bigl( {\Delta^2 \over q^2} \Bigr)^2 B_2(\lambda,
\alpha, \xi, g^2) + \Bigl( {\Delta^2 \over q^2} \Bigr)^3
B_3(\lambda, \alpha, \xi, g^2) + f_3(q^2) + ....,
\end{equation}
where we used notations (5.2), since the coefficients of the
above-used Taylor expansions depend in general on the same set of
parameters: $\lambda, \alpha, \xi, g^2$. The invariant function
$f_3(q^2)$ is dimensionless and regular at small $q^2$; otherwise
it remains arbitrary. The generalization on the next iterations is
almost obvious. Let us only note that in this case more terms in
the corresponding Taylor expansions should be kept "alive".

\subsection{The exact structure of the general iteration solution}

Substituting the generalization of the expansion (5.17) on all
iterations and omitting the tedious algebra, the general iteration
solution of the gluon SD equation (3.22) for the regularized full
gluon propagator (3.16) can be exactly decomposed as the sum of
the two principally different terms as follows:

\begin{eqnarray}
D_{\mu\nu}(q; \Delta^2) = D^{INP}_{\mu\nu}(q; \Delta^2)+
D^{PT}_{\mu\nu}(q) &=& i T_{\mu\nu}(q) {\Delta^2 \over (q^2)^2}
\sum_{k=0}^{\infty} (\Delta^2 / q^2)^k \sum_{m=0}^{\infty}
\Phi_{k,m}(\lambda, \alpha,
\xi, g^2) \nonumber\\
&+& i \Bigr[ T_{\mu\nu}(q) \sum_{m=0}^{\infty} A_m(q^2) + \xi
L_{\mu\nu}(q) \Bigl] {1 \over q^2},
\end{eqnarray}
where the superscript "INP" stands for the intrinsically NP part
of the full gluon propagator. We distinguish between the two terms
in Eq. (5.18) by the character of the corresponding IR
singularities and the explicit presence of the mass gap (see
below). Let us emphasize that the general problem of convergence
of the formally regularized series (5.18) is irrelevant here.
Anyway, the problem how to remove all types of the UV divergences
(overlapping \cite{14} (see some remarks below as well) and
overall \cite{1,2,6,7,8,9}) is a standard one. Our problem will be
how to deal with severe IR singularities due to their novelty and
genuine NP character. Fortunately, there already exists a
well-elaborated mathematical formalism for this purpose, namely
the distribution theory (DT) \cite{15} to which the DRM \cite{11}
should be correctly implemented (see also Refs. \cite{10,16}).

The INP part of the full gluon propagator is characterized by the
presence of severe power-type (or equivalently NP) IR
singularities $(q^2)^{-2-k}, \ k=0,1,2,3,...$. So these IR
singularities are defined as more singular than the power-type IR
singularity of the free gluon propagator $(q^2)^{-1}$, which thus
can be defined as the PT IR singularity. The INP part depends only
on the transversal degrees of freedom of gauge bosons. Though its
coefficients $\Phi_{k,m}(\lambda, \alpha, \xi, g^2)$ may
explicitly depend on the gauge-fixing parameter $\xi$, the
structure of this expansion itself does not depend on it. It
vanishes as the mass gap goes formally to zero, while the PT part
survives. The INP part of the full gluon propagator in Eq. (5.18)
is nothing but the corresponding Laurent expansion in integer
powers of $q^2$ accompanied by the corresponding powers of the
mass gap squared and multiplied by the sum over the
$q^2$-independent factors, the so-called residues $\Phi_k(\lambda,
\alpha, \xi, g^2) = \sum_{m=0}^{\infty} \Phi_{k,m}(\lambda,
\alpha, \xi, g^2)$. The sum over $m$ indicates that an infinite
number of iterations (all iterations) of the corresponding
regularized skeleton loop integrals invokes each severe IR
singularity labelled by $k$. It is worth emphasizing that now this
Laurent expansion cannot be summed up into anything similar to the
initial Eq. (3.17), since its residues at poles gain additional
contributions due to the splitting/shifting procedure, i.e., they
become arbitrary. However, this arbitrariness is not a problem,
because severe IR singularities should be treated by the DRM
correctly implemented into the DT. For this the dependence of the
residues on their arguments is all that matters and not their
concrete values. The PT part of the full gluon propagator, which
has only the PT IR singularity, remains undetermined. In the PT
part the sum over $m$ again indicates that all iterations
contribute to the PT gluon form factor $d^{PT}(q^2) =
\sum_{m=0}^{\infty} A_m(q^2)$. What we know about $A_m(q^2)$
functions is only that they are regular functions at small $q^2$;
otherwise remaining arbitrary but $d^{PT}(q^2)$ should satisfy AF
at large $q^2$. This is the price we have paid to fix exactly the
functional dependence of the INP part of the full gluon
propagator. Just this part gives rise to the dominant
contributions to the numerical values of physical quantities in
low-energy QCD (see below as well). In Refs. \cite{10,16,17} we
came to the same structure (5.18) but in a rather different way.

Both terms in Eq. (5.18) are valid in the whole energy/momentum
range, i.e., they are not asymptotics. At the same time, we have
achieved the exact separation between the two terms responsible
for the NP (dominating in the IR ($q^2 \rightarrow 0$)) and the
nontrivial PT (dominating in the UV ($q^2 \rightarrow \infty$))
dynamics in the true QCD vacuum. It is worth emphasizing once more
that we exactly distinguish between the two terms in Eq. (5.18) by
the character of the corresponding IR singularities. This first
necessary condition includes the existence of a special
regularization expansion for severe (i.e., NP) IR singularities,
while for the PT IR singularity it does not exist \cite{10,15}.
The second sufficient condition is the explicit presence of the
mass gap (when it formally goes to zero then the PT phase survives
only). So the above-mentioned separation is not only exact but
unique as well. Evidently, it is only possible on the basis of the
corresponding decomposition of the full gluon form factor in Eq.
(3.16) as follows:

\begin{equation}
d(q^2) = d(q^2) - d^{PT}(q^2)+ d^{PT}(q^2)= d^{INP} (q^2) +
d^{PT}(q^2).
\end{equation}
As explained above this separation is exact and unique within the
general iteration solution. Due to the character of the IR
singularity the longitudinal component of the full gluon
propagator should be included into its PT part, so its INP part
becomes automatically transversal.

In summary, the general iteration solution (5.18) is inevitably
severely singular in the IR limit ($q^2 \rightarrow 0$), and this
does not depend on the special gauge choice (see discussion below
as well).

\subsection{Remarks on overlapping divergences}

The mass gap which appears first in the gluon SD equation (3.13)
is the main object we have worried about to demonstrate explicitly
its crucial role within our approach. Let us make, however, a few
remarks in advance. As it follows from the standard gluon SD
equation (3.13), the corresponding equation for the gluon
self-energy looks like

\begin{equation}
D^{-1}(q) = D^{-1}_0(q) - q^2 \Pi(q^2; D) - \Delta^2(\lambda; D),
\end{equation}
where we omit the tensor indices as well as the longitudinal part
of the subtracted gluon self-energy, for simplicity. In order to
unravel overlapping UV divergence problems in QCD, the necessary
number of the differentiation with respect to the external
momentum should be done first (in order to lower divergences).
Then the point-like vertices, which are present in the
corresponding skeleton loop integrals should be replaced by their
full counterparts via the corresponding integral equations.
Finally, one obtains the corresponding SD equations which are much
more complicated than the standards ones, containing different
scattering amplitudes, which skeleton expansions are, however,
free from the above-mentioned overlapping divergences. Of course,
the real procedure \cite{14} (and references therein) is much more
tedious than briefly described above. However, even at this level,
it is clear that by taking derivatives with respect to the
external momentum $q$ in the SD equation for the gluon self-energy
(5.20), the main initial information on the mass gap will be
totally lost. Whether it will be somehow restored or not at the
later stages of the renormalization program is not clear at all.
Thus in order to remove overlapping UV divergences ("the water")
from the SD equations and skeleton expansions, we are in danger to
completely loose the information on the dynamical source of the
mass gap ("the baby") within our approach. In order to avoid this
danger and to be guaranteed that no any dynamical information are
lost, we are using the standard gluon SD equation (3.13). The
presence of any kind of UV divergences (overlapping and usual
(overall)) in the skeleton expansions will not cause any problems
in order to detect the mass gap responsible for the IR structure
of the true QCD vacuum. In other words, the direct iteration
solution of the standard gluon SD equation (3.13) or equivalently
(3.22) is reliable to realize a mass gap, and thus to make its
existence perfectly clear. The problem of convergence of such
regularized skeleton loop series which appear in Eq. (5.18) is
completely irrelevant in the context of the present investigation.
Anyway, we keep any kind of UV divergences under control within
our method, since we are working with the regularized quantities.
At the same time, the existence of a mass gap responsible for the
IR structure of the full gluon propagator does not depend on
whether overlapping divergences are present or not in the SD
equations and corresponding skeleton expansions. All this is the
main reason why our starting point is the standard gluon SD
equation (3.13) for the unrenormalized Green's functions (this
also simplifies notations). See discussion below as well in order
to understand why the problem of overlapping divergences is not
important for us. Roughly speaking, if one starts from the UV
renormalized equations from the very beginning then the
information about mass gap will be totally lost. So, one should
start from the unrenormalized equation (but for the regularized
quantities), then to release a mass gap as it was described. The
next step is to perform the IR renormalized program within the
general iteration solution, and on the last step to perform the UV
renormalization program.

\section{Discussion}

It is worth recalling now that in the NP approach to QCD the
ghosts are already not sufficient to guarantee the cancellation of
unphysical degrees of freedom of gauge bosons. The standard way to
make the full gluon propagator purely transversal is to choose the
Landau gauge $\xi=0$ from the very beginning. The system of the SD
equations and the corresponding Green's functions, which should
satisfy them, is explicitly gauge-dependent. So in principle to
choose gauge by hand at this level should not be a problem. The
only request is that the $S$-matrix elements, describing the
corresponding quantities and processes in low-energy QCD, should
not depend explicitly on the gauge choice. However, as a subject
for discussion let us formulate here a general method how to make
the gluon propagator relevant for NP QCD to be automatically
transversal.

\subsection{The necessity of the subtractions}

Many important quantities in QCD, such as the gluon and quark
condensates, the topological susceptibility, the Bag constant
(which is just the difference between the PT and NP vacuum energy
densities, see below an example 2), etc., are defined only beyond
the PT \cite{18,19,20}. This means that they are determined by
such $S$-matrix elements (correlation functions) from which all
types of the PT contributions should be, by definition,
subtracted.

It is worth emphasizing that such type of the subtractions are
inevitable also for the sake of self-consistency. In low-energy
QCD there exist relations between different correlation functions,
for example, the Witten-Veneziano (WV) and Gell-Mann-Oakes-Renner
(GMOR) formulae. The former \cite{21,22} relates the pion decay
constant and the mass of the $\eta'$ meson to the topological
susceptibility. The latter \cite{19,23} relates the chiral quark
condensate to the pion decay constant and its mass. The famous
trace anomaly relation (see, for example Refs. \cite{19,22} and
references therein) relates the Bag constant (which is the truly
NP vacuum energy density, apart from the sign) to the gluon and
quark condensates. Defining thus the topological susceptibility
and the gluon and quark condensates by the subtraction of all
types of the PT contributions, it would not be self-consistent to
retain them in the correlation function, determining the pion
decay constant, and in the expressions for the pion and $\eta'$
meson masses.

A few additional remarks about the subtraction of the PT
contributions are in order. Let us remind that in lattice QCD
\cite{1,2,24} such kind of an equivalent procedure also exists. In
order to prepare an ensemble of lattice configurations for the
calculation of any NP quantity or to investigate some NP
phenomena, the excitations and fluctuations of gluon fields of the
PT origin and magnitude should be "washed out" from the vacuum.
This goal is usually achieved by using "Perfect Actions",
"cooling", "cycling", etc., (see, for example, Refs. \cite{3,4}
and references therein). Evidently, this is very similar to our
method in continuous QCD (for details see below).

From QCD sum rules \cite{19} it is well known that AF is stopped
by power-type terms reflecting the growth of the coupling in the
IR. Approaching the deep IR region from above, the IR sensitive
contributions were parameterized in terms of a few quantities (the
gluon and quark condensates, etc.), while the direct access to NP
effects (i.e., to the deep IR region) was blocked by the IR
divergences \cite{19,25}. In order to calculate the gluon
condensate the corresponding subtraction of the PT gluon
propagator integrated out over the deep IR region (where it
certainly fails) should be also done (see discussion given by
Shifman in Ref. \cite{4}). In order to correctly calculate the
gluon condensate by analytic methods the necessity of the
subtraction of the PT part of the effective coupling constant
(integrated out) has been explicitly shown in recent papers
\cite{26,27} as well.

There also exists very serious argument in favor of inevitability
of the above-discussed subtractions of the PT contributions at all
levels and all types in order to fix the gauge of truly NP QCD. In
his pioneering paper \cite{28} Gribov has investigated the
quantization problem of non-Abelian gauge theories using the
functional integral representation of the generating functional
for non-Abelian gauge fields. It has been explicitly shown that
the standard Fadeev-Popov (FP) prescription fails to fix the gauge
uniquely and therefore should be modified, i.e., it is not enough
to eliminate arbitrary degrees of freedom from the theory. In
other words, there is an ambiguity in the gauge-fixing of
non-Abelian gauge fields (the so-called Gribov ambiguity
(uncertainty), which results in Gribov copies and vice versa). To
resolve this problem Gribov has explicitly demonstrated that the
modification reduces simply to an additional limitation on the
integration range in the functional space of non-Abelian gauge
fields, which consists in integrating only over the fields for
which the FP determinant is positive \cite{28} (introducing thus
the so-called Gribov horizon in the functional space, see also
Ref. \cite{29}). As emphasized by Gribov, this affects the IR
singularities of the PT and results in a linear increase of the
charge interaction at large distances (see also remarks and
discussion below in subsection D).

\subsection{Restoration of the transversality of gauge bosons}

Anyway, to calculate correctly any truly NP quantity in low-energy
QCD from first principles one has to begin with making
subtractions at the fundamental quark-gluon level. First of all,
it is necessary to fix a scale responsible for the NP dynamics in
the system. The second step is to set it to zero in order to
recover the corresponding PT phase in the system. In our case for
the NP gluon propagator the formal PT limit is $\Delta^2 =0$. So
from Eqs. (3.16) and (3.17) the PT gluon propagator becomes

\begin{equation}
D^{PT}_{\mu\nu}(q) = i \left\{ T_{\mu\nu}(q) d^{PT}(q^2) + \xi
L_{\mu\nu}(q) \right\} {1 \over q^2},
\end{equation}
where

\begin{equation}
d^{PT}(q^2) = {1 \over 1 + \Pi(q^2; D^{PT})},
\end{equation}
and from the gluon SD equation (3.22) in this limit, one recovers
the corresponding gluon SD equation as follows:

\begin{equation}
D^{PT}_{\mu\nu}(q) = D^0_{\mu\nu}(q) - T_{\mu\sigma}(q) \Pi(q^2;
D^{PT}) D^{PT}_{\sigma\nu}(q),
\end{equation}
which, of course, coincides with Eq. (3.1), since
$\Pi^s_{\rho\sigma}(q) = \Pi_{\rho\sigma}(q)$ in this limit (see
Eq. (3.7)).

The truly NP gluon propagator is to be defined as follows:

\begin{equation}
D^{NP}_{\mu\nu}(q; \Delta^2) = D_{\mu\nu}(q; \Delta^2)-
D_{\mu\nu}(q; \Delta^2=0) = D_{\mu\nu}(q; \Delta^2)-
D^{PT}_{\mu\nu}(q),
\end{equation}
so that the full gluon propagator becomes an exact sum of the two
different terms

\begin{equation}
D_{\mu\nu}(q; \Delta^2) = D^{NP}_{\mu\nu}(q; \Delta^2) +
D^{PT}_{\mu\nu}(q).
\end{equation}
The principal difference between the full gluon propagator
$D_{\mu\nu}(q; \Delta^2)$ and the truly NP gluon propagator
$D^{NP}_{\mu\nu}(q; \Delta^2)$ is that the latter one is free from
the PT "contaminations", while the former one, being also NP, is
"contaminated" by the PT contributions. Since the PT limit is
uniquely defined in our approach, the separation between the truly
NP and PT gluon propagators is uniquely defined as well. So from
Eq. (6.5) it follows that when the mass gap, which is responsible
for the NP dynamics, is set to zero $\Delta^2=0$, then only the PT
phase survives, i.e., the full gluon propagator is reduced to its
nontrivial PT counterpart. That is a reason why we call
$\Delta^2=0$ the PT limit.

Both terms in the exact decomposition (6.5) are valid in the whole
energy/momentum region, i.e., they are not asymptotics. At the
same time, we achieved the exact separation between the two terms
responsible for the NP (dominating in the IR ($q^2 \rightarrow
0$)) and the nontrivial PT (dominating in the UV ($q^2 \rightarrow
\infty$)) dynamics in the true QCD vacuum. This is so indeed
because the PT limit $\Delta^2 =0$ is equivalent to the UV limit
$q^2 \rightarrow \infty$ (see Eq. (3.17)), which means in its turn
that the PT limit is in agreement with AF within our approach.
Thus if in any model gluon propagator a scale responsible for the
NP dynamics cannot be fixed explicitly, then in order to recover
the truly NP part its behavior at infinity should be subtracted.
In this case, however, the separation between the NP and PT phases
may not be unique. Also, we distinguish between the two different
phases in QCD not by the strength of the coupling constant, but by
the presence of a mass gap (in this case the coupling constant
plays no any role as it follows from our consideration).

From the definition (6.4) and all the above displayed relations,
the truly NP gluon propagator finally becomes

\begin{equation}
D^{NP}_{\mu\nu}(q; \Delta^2) = i T_{\mu\nu}(q) d^{NP}(q^2;
\Delta^2) {1 \over q^2},
\end{equation}
where the truly NP gluon form factor is

\begin{equation}
d^{NP}(q^2; \Delta^2) = {\Pi(q^2; D^{PT}) - \Pi(q^2; D) -
{\Delta^2(\lambda; D) \over q^2} \over [1 + \Pi(q^2; D) +
{\Delta^2(\lambda; D) \over q^2}] [1 + \Pi(q^2; D^{PT}]},
\end{equation}
and it may be treated as the truly NP effective charge as well.
Evidently, the truly NP gluon propagator is manifestly
transversal, i.e., does not depend explicitly on the gauge-fixing
parameter. This results in the fact that both the full gluon
propagator (3.16) and the PT gluon propagator (6.1) satisfy the
color gauge invariance condition (3.15) within our approach.
Otherwise the truly NP gluon propagator cannot not be really
transversal. Let us repeat once more that Eq. (6.7) is not a
solution, but it displays a general structure of the truly NP
gluon form factor. To establish a possible type of solution,
however, a much more convenient starting point is the full gluon
form factor (see Eq. (3.17)). It will give a possible type of
solution for Eq. (6.7) as well. In the PT limit $\Delta^2 =0$ the
full gluon propagator coincides with its nontrivial PT part, so
the ghosts can now fulfil their role to cancel unphysical degrees
of freedom of gauge bosons, since the transversality of the gluon
self-energy is to be restored in this limit (see Eqs. (3.7) and
(3.11)).

In summary, our prescription as how to guarantee the
transversality of the gluon propagator relevant for calculations
of the truly NP physical quantities and processes from first
principles in low-energy QCD is to be briefly formulated as
follows:

(i). The corresponding scale responsible for the NP dynamics (the
mass gap) is to be fixed. The dependence of the full gluon
propagator on the mass gap should be only regular. If the
dependence of the full gluon propagator on a some scale parameter
is singular, then it cannot be chosen as the mass gap. In
particular, this means that the invariant function $\Pi(q^2; D)$
depends (if any) on the mass gap only regularly.

(ii). The exact decomposition (6.5) should be provided.

(iii). All terms, containing $D^{PT}_{\mu\nu}(q)$, should be
discarded from all the relations, equations, etc. i.e., to put
$D_{\mu\nu}(q; \Delta^2) = D^{NP}_{\mu\nu}(q; \Delta^2)$
everywhere. Let us note in advance that in the case of the general
iteration solution the full gluon propagator should be replaced by
its INP part, i.e., $D_{\mu\nu}(q; \Delta^2) = D^{INP}_{\mu\nu}(q;
\Delta^2)$ (see below).

\subsection{The truly NP massive solution}

In accordance with our method how to define the truly NP gluon
propagator described above, we have to establish first the scale
responsible for the NP dynamics in the system under consideration
and then to set it to zero. For the massive-type solution (4.6)
the gluon mass may serve as a scale responsible for its NP
dynamics, since in the PT limit $\Delta^2(\lambda, \xi) =0$, the
effective gluon mass is also zero, $m_g^2(\lambda, \xi) =0$. Then
the nontrivial PT part of the full gluon propagator becomes

\begin{equation}
D^{PT}_{\mu\nu}(q) = i T_{\mu\nu}(q) {Z^{PT}_3 \over q^2 [ 1 +
\Pi^R(q^2; 0)]} + i \xi L_{\mu\nu}(q) {1 \over q^2}.
\end{equation}
Here the PT gluon propagator pole is going to zero and the gluon
renormalization constant becomes a standard one, namely

\begin{equation}
Z^{PT}_3 = [1 + \Pi(0; \xi)]^{-1}.
\end{equation}

The truly NP gluon propagator defined in Eq. (6.4) as the
difference between the full gluon propagator (4.6) and the PT
gluon propagator (4.8) becomes

\begin{equation}
D^{NP}_{\mu\nu}(q; m^2_g) = i T_{\mu\nu}(q) \Bigl[ {Z_3 \over (q^2
+ m^2_g) [ 1 + \Pi^R(q^2; m^2_g)]} - {Z^{PT}_3 \over q^2[ 1 +
\Pi^R(q^2; 0)]} \Bigr].
\end{equation}
In principle, this type of solution demonstrates the propagation
of the two different purely transversal gluons: massive, i.e., the
NP (first term) and massless, i.e., the nontrivial PT (second
term) ones. Concluding, let us make one speculative remark. Such
structure of the purely transversal gluon propagator can be a hint
of a similar mechanism of a mass creating without still missing
Higgs particle in Standard model. So massless gluons may
correspond to the photons, while massive solutions (let us remind
that there is also a second solution, see section IV) may
correspond to $Z$ and $W^{\pm}$ bosons.

\subsection{The INP iteration solution}

In the general iteration solution (5.18) we distinguish first of
all between its two parts by the character of the IR singularities
(see above). This means the automatical presence of a mass gap,
since any deviation of the behavior of the full gluon propagator
from the free one requires the existence of a mass scale
parameter. Thus for this solution only the explicit presence of a
mass gap becomes the second sufficient condition to separate the
different terms in the full gluon propagator. For all other
possible types of solutions (for example, for the above-described
massive-type solution) the presence of a mass gap is necessary and
sufficient for this purpose. Making the subtraction in accordance
with this exact criterion in Eq. (5.18), one concludes that the
role of the gluon propagator responsible for the NP dynamics
should be assigned to its INP part, which becomes truly NP,
indeed, i.e., it vanishes in the PT limit in accordance with the
general prescription. Let us recall that the separation between
the INP and PT parts in the full gluon propagator is unique as
well. It is based on the existence of a special regularization
expansion for severe (i.e., NP) IR singularities in the DT
complemented by the DRM. For the PT IR singularity such kind of
the expansion does not exist \cite{10,15}. This emphasizes the
special character of the nonlinear iteration solution. Thus, one
obtains

\begin{equation}
D^{INP}_{\mu\nu}(q; \Delta^2) = i T_{\mu\nu}(q) d^{INP}(q^2;
\Delta^2) {1 \over q^2} = i T_{\mu\nu}(q) {\Delta^2 \over (q^2)^2}
\sum_{k=0}^{\infty} (\Delta^2 / q^2)^k \Phi_k(\lambda, \alpha,
\xi, g^2),
\end{equation}
where

\begin{equation}
\Phi_k(\lambda, \alpha, \xi, g^2) = \sum_{m=0}^{\infty}
\Phi_{k,m}(\lambda, \alpha, \xi, g^2),
\end{equation}
and the effective charge in this case is to be defined as follows:

\begin{equation}
d^{INP}(q^2; \Delta^2) \equiv \alpha^{INP}_s(q^2; \Delta^2) =
{\Delta^2 \over q^2} \sum_{k=0}^{\infty} (\Delta^2 / q^2)^k
\Phi_k(\lambda, \alpha, \xi, g^2).
\end{equation}
Evidently, The INP part depends only on the transversal degrees of
freedom of gauge bosons as it is required, by definition. Also,
its functional dependence is uniquely fixed up to the expressions
for the residues $\Phi_k(\lambda, \alpha, \xi, g^2)$, and it is
valid in the whole energy/momentum range. At large momentum it
looks formally as an Operator Product Expansion (OPE) of the gluon
propagator. However, it is completely suppressed in this limit
($q^2 \rightarrow \infty$) in comparison with the PT part of the
full gluon propagator. As underlined above, what we worried about
is its behavior in the IR limit ($q^2 \rightarrow 0$), which can
be correspondingly treated within the DRM correctly implemented
into the DT. Only this investigation will allow one to deduce
whether the severe IR structure of the gluon propagator survives
or not. Up to this moment, however, the INP part is a Laurent
expansion in powers of $q^2$, multiplied by the corresponding
powers of the mass gap, and it starts from the simplest severe (or
equivalently NP) power-type IR singularity which is only one
possible in four-dimensional QCD, namely $(q^2)^{-2}$
\cite{10,15}. Let us also remind that no approximation/truncations
are made for the corresponding regularized constant skeleton loop
integrals, contributing to the residues $\Phi_k(\lambda, \alpha,
\xi, g^2)$ over all iterations.

The unavoidable existence of the INP part of the full gluon
propagator within its general iteration solution (5.18) makes the
principal difference between non-Abelian QCD and Abelian QED,
where such kind of term in the full photon propagator is certainly
absent (in the former theory there is direct coupling between
massless gluons which finally leads to the dynamical generation of
a mass gap, while in the latter one there is no direct coupling
between massless photons). Precisely this term violets the cluster
properties of the Wightman functions \cite{30}, and thus validates
the Strocchi theorem \cite{31}, which allows for such IR singular
behavior of the full gluon propagator. Contrary to QCD, the full
photon propagator may have only the PT-type IR singularity (see
Eq. (2.8) above).

Though the residues at poles shown in Eq. (6.2) may explicitly
depend on the gauge-fixing parameter $\xi$, the zero momentum
modes enhancement (ZMME) effect itself (represented in the INP
part) does not depend on it. Since we do not specify explicitly
the value of the gauge-fixing parameter $\xi$, the ZMME effect
takes place at its any value. This is very similar to AF. It is
well known that the exponent which determines the logarithmic
deviation of the full gluon propagator from the free one in the UV
region ($q^2 \gg \Lambda^2_{QCD}$) explicitly depends on the
gauge-fixing parameter. At the same time, AF itself does not
depend on it, i.e., it takes place at any $\xi$.

The QCD Lagrangian does not contain a mass gap. However, we
discovered that the mass scale parameter responsible for the NP
dynamics in the IR region should exist in the true QCD ground
state. At the level of the gluon SD equation it is hidden in the
skeleton loop contributions into the gluon self-energy. Within the
general iteration solution it explicitly shows up (and hence the
corresponding severe IR singularities) when the gluon momentum
goes to zero. At the fundamental quark-gluon (i.e., Lagrangian)
level the main dynamical source of a mass gap is the
self-interaction of massless gluons, i.e., the NL dynamics of QCD.
The triple gluon vertex vanishes when all the gluon momenta
involved go to zero ($T_3(0,0) =0$), while its four-gluon
counterpart survives ($T_4(0,0,0) \neq 0$). Then one may think
that the latter one plays more important role than the former one
in the IR structure of the gluon SD equation and thus in the
arising of the mass gap mainly from quartic gluon potential
(Feynman \cite{5,32} has also arrived at the same conclusion but
on a different basis). The skeleton tadpole term

\begin{equation}
\Pi_t(D) =  g^2 \int {i d^4 q_1 \over (2 \pi)^4} T^0_4 (q_1, 0,0,
-q_1)D(q_1),
\end{equation}
(for simplicity, we omit the tensor and color indices), which
explicitly violates the transversality of the gluon self-energy
(see Eqs. (3.5) and (3.6)), contains only the four-gluon vertex.
So there is no doubt in the important role of quartic gluon
potential in NP QCD, indeed.

Thus the true QCD vacuum is really beset with severe IR
singularities. Within the general iteration solution they should
be summarized (accumulated) into the full gluon propagator and
effectively correctly described by its structure in the deep IR
domain, exactly represented by its INP part. It is worth
emphasizing here that due to the arbitrariness of the
above-mentioned residues $\Phi_k(\lambda, \alpha, \xi, g^2)$ at
poles $(q^2)^{-2-k}, \ k=0,1,2,3,...$, there is no smooth gluon
propagator within the general iteration solution. The second step
is to assign a mathematical meaning to the integrals, where such
kind of severe IR singularities will explicitly appear, i.e., to
define them correctly in the IR region \cite{10,15}. Just this IR
violent behavior makes QCD as a whole an IR unstable theory, and
therefore it may have no IR stable fixed point, indeed \cite{1},
which means that QCD itself might be a confining theory without
involving some extra degrees of freedom
\cite{33,34,35,36,37,38,39,40}.

The INP part of the full gluon propagator (6.11) depends only on
the transversal ("physical") degrees of freedom of gauge bosons,
by construction, due to the above-described in detail subtraction
procedure at the gluon propagator level. All the problems with the
gauge-fixing discovered by Gribov \cite{28} in the functional
space should be attributed to the PT part of the gluon propagator
within its general iteration solution. Making the above-mentioned
subtraction, in order to proceed to the gluon propagator relevant
for NP QCD, we thus will make it free of the gauge-fixing
ambiguity in the momentum space (the implicit dependence of the
residues on the gauge-fixing parameter is not dangerous). This
once more emphasizes the necessity and importance of making
subtraction in order to make the theory at the fundamental
quark-gluon level free of this problem, which otherwise will
plague the dynamics of any essentially NL gauge systems \cite{29}.

\subsubsection{An example 1}

It is instructive to describe the procedure of the PT subtractions
of all types in more detail within the general iteration solution.
For example, the gluon condensate can be formally defined (up to
some unimportant for our purpose numerical factors) as the
effective coupling integrated out, namely

\begin{equation}
\langle{0} | G^2 | {0}\rangle \sim \int_0^{\infty} \alpha_s(q^2)
q^2 dq^2.
\end{equation}
In accordance with Eq. (5.19) the effective charge is to be
identically decomposed as follows: $\alpha_s(q^2)= \alpha_s(q^2) -
\alpha_s^{PT}(q^2) + \alpha_s^{PT}(q^2)= \alpha_s^{INP}(q^2)+
\alpha_s^{PT}(q^2)$. Let us introduce further the effective scale
$q^2_{eff}$, which separates the NP region from the PT one (it
looks something like the above-mentioned Gribov horizon, but in
the much simpler momentum space). Then the initial integral
becomes the sum of four terms

\begin{eqnarray}
\langle{0} | G^2 | {0}\rangle \sim \int_0^{\infty} \alpha_s(q^2)
q^2 dq^2 &=& \int_0^{q_{eff}^2} \alpha_s^{INP}(q^2) q^2 dq^2 +
\int_{q^2_{eff}}^{\infty} \alpha_s^{INP}(q^2) q^2 dq^2
\nonumber\\
&+& \int_0^{q^2_{eff}} \alpha_s^{PT}(q^2) q^2 dq^2 +
\int_{q_{eff}^2}^{\infty} \alpha_s^{PT}(q^2) q^2 dq^2.
\end{eqnarray}
Within our approach the NP region includes not only the deep IR
one but even more than that (it may includes some substantial part
of the intermediate region as well). In the INP QCD all the three
last integrals reproduce the different types of the PT
contributions despite some of them might be finite numbers,
nevertheless. So in our theory the "physical" gluon condensate is
to be defined by subtracting all these integrals from the initial
one, i.e., one has to put

\begin{eqnarray}
 \langle{0} | G^2 | {0}\rangle_{"ph"} & \sim & \int_0^{\infty}
\alpha_s(q^2) q^2 dq^2 - \int_{q^2_{eff}}^{\infty}
\alpha_s^{INP}(q^2) q^2 dq^2 - \int_0^{q^2_{eff}}
\alpha_s^{PT}(q^2) q^2 dq^2 - \int_{q_{eff}^2}^{\infty}
\alpha_s^{PT}(q^2) q^2 dq^2
\nonumber\\
&=& \int_0^{q_{eff}^2} \alpha_s^{INP}(q^2) q^2 dq^2,
\end{eqnarray}
and thus it is free of all types of the PT contributions, indeed,
since the right-hand-side of this equation is the INP effective
charge integrated out over the NP region ($0 \leq q^2 \leq
q_{eff}^2$). Of course, this expression cannot be used for actual
calculation of the gluon condensate, since the INP effective
charge (6.13) is not yet IR and UV renormalized. However, for the
preliminary actual calculations of the gluon condensate free of
all types of the PT contributions and not using a weak coupling
limit solution to the $\beta$ function see our papers \cite{18,26}
(and references therein).

In QCD sum rules the INP effective charge $\alpha_s^{INP}(q^2)$ is
not known. So omitting it in Eq. (6.16), one obtains

\begin{equation}
\langle{0} | G^2 | {0}\rangle \approx \int_0^{\infty}
\alpha_s(q^2) q^2 dq^2 - \int_0^{q^2_{eff}} \alpha_s^{PT}(q^2) q^2
dq^2 - \int_{q_{eff}^2}^{\infty} \alpha_s^{PT}(q^2) q^2 dq^2.
\end{equation}
Omitting further the second integral (see the discussion given by
Shifman in Ref. \cite{4}), which is nothing else but the PT
effective charge integrated out over the NP region, one finally
obtains

\begin{equation}
\langle{0} | G^2 | {0}\rangle_{"ph"} \approx \int_0^{\infty}
\alpha_s(q^2) q^2 dq^2 - \int_{q_{eff}^2}^{\infty}
\alpha_s^{PT}(q^2) q^2 dq^2,
\end{equation}
i.e., in this theory the gluon condensate is again the difference
between an infinite initial integral and infinite PT tail
(evidently, in this integral it is justified to approximate the
full effective charge by its PT counterpart, indeed). So that the
above-mentioned difference is finite, and in fact it is a rather
good approximation to the exact definition of the gluon condensate
within our approach in Eq. (6.17), i.e., to its right-hand-side.
Within this formalism the only problem here is the point of the
subtraction $q^2_{eff}$ since the separation "hard vs soft" gluon
momenta in this theory is not exact, while in our theory it is
exactly fixed through the mass gap. Nevertheless, our and QCD sum
rules values for the gluon condensate are rather close to each
other \cite{26}. Especially very good agreement is achieving when
our value is recalculated at the $1 \ GeV$ scale, i.e., when we
put $q^2_{eff} = 1 \ GeV^2$. It is very reasonable value for the
effective scale $q^2_{eff}$ separating the NP region from the PT
one.

\subsubsection{An example 2}

In close connection with the gluon condensate is one of the main
characteristics of the true QCD ground state is the
above-mentioned Bag constant. It is just defined as the difference
between the PT and NP vacuum energy densities (VED). So, we can
symbolically put $B = VED^{PT} - VED$, where $VED$ is the NP but
"contaminated" by the PT contributions (i.e., this is a full $VED$
like the full gluon propagator). At the same time, in accordance
with our method we can continue as follows: $B = VED^{PT} - VED =
VED^{PT} - [VED - VED^{PT} + VED^{PT}] = VED^{PT} - [VED^{INP} +
VED^{PT}] = - VED^{INP} > 0$, since the VED is always negative.
Thus the Bag constant is nothing but the INP VED, apart from the
sign, by definition, and thus is completely free of the PT
"contaminations". For how to correctly define and actually
calculate the Bag constant from first principles by making all
necessary subtractions at all level see again our paper \cite{26},
where the relation between the Bag constant and gluon condensate
is explicitly shown as well.

\subsubsection{An example 3}

The chiral quark condensate is formally defined as follows
(Euclidean signature):

\begin{equation}
<0|\bar{q} q|0>_0 = \int {id^4p \over (2 \pi)^4} Tr S(p),
\end{equation}
where the trace over color and Dirac matrices is understood. Here
$S(p)= i[\hat p A(p^2) + B(p^2)]$ is the full quark propagator.
After trivial derivation it becomes

\begin{equation}
<0|\bar{q} q|0>_0 \sim - \int_0^{\infty} p^2 dp^2 B(p^2),
\end{equation}
where and below we will omit all numerical factors as unimportant
for our purpose. In accordance with our method the "physical"
chiral quark condensate should be defined as follows:

\begin{equation}
<0|\bar{q} q|0>_0^{"ph"} \sim - \int_0^{\infty} p^2 dp^2 B(p^2) +
\int_{q^2_{eff}}^{\infty} p^2 dp^2 B(p^2) =  - \int_0^{q^2_{eff}}
p^2 dp^2 B(p^2),
\end{equation}
where the effective scale $q^2_{eff}$ numerically, in principle,
differs from the Yang-Mills (YM) effective scale introduced in the
previous examples. However, its chosen value $q^2_{eff} = 1 \
GeV^2$ might be a good approximation for full QCD as well. The
only thing remaining to do is to substitute for the quark running
mass function $B(p^2)$ solution of the quark SD equation in the
chiral limit based on the INP gluon propagator (6.11). For actual
preliminary calculations of the chiral quark condensate and other
chiral QCD parameters within the INP approach to QCD see our
papers \cite{41,42}.

In summary, our consideration in general and these symbolic
examples within the INP solution to QCD in particular clearly
shows how to correctly calculate the physical observables (and
related quantities) from first principles in low-energy QCD. The
subtractions of the PT contributions of all types and at all
levels are necessary to be made for this purpose. This means that
INP QCD is the UV finite theory, by definition, though both
renormalization programs are still needed in order to correctly
define the corresponding Green's functions and their solutions
\cite{10}.

\section{Conclusions}

Our consideration at this stage is necessarily formal, since the
mass gap $\Delta^2$ remains neither IR nor UV renormalized yet. At
this stage it has been only regularized, i.e., $\Delta^2 \equiv
\Delta^2(\lambda, \alpha, \xi, g^2)$. However, there is no doubt
that it will survive both multiplicative renormalization (MR)
programs (which include the corresponding removal of both
$\lambda$ and $\alpha$ parameters). The UVMR program is not our
problem (it is a standard one \cite{1,2,6,7,8,9,14}, and anyway,
as underlined above, it should follow after IRMR one is
performed). Within the INP solution to QCD our problem is the IRMR
program in order to render the whole theory finite, i.e., to make
it free from all types of severe IR singularities parameterized in
terms of the IR regularization parameter as it goes to zero at the
final stage. It is not a simple task due to its novelty and really
NP character. It requires much more tedious technical work how to
correctly implement the DRM into the DT, and it is left to be done
elsewhere (for some preliminary aspects of the IRMR program see
Ref. \cite{10}).

It is worth noting  that the mass gap which appears in the gluon
SD equation cannot be in principle the same one which appears in
the INP part of the general iteration solution, though we have
identified them, for simplicity. Let us denote the renormalized
version of the mass gap $\Delta^2(\lambda; D)$ (i.e., which
appears in the gluon SD equation) as $\Delta^2_{JW}$ and call it
the Jaffe-Witten (JW) mass gap \cite{5} (see theorem below as
well). At the same time, the renormalized version of our mass gap
$\Delta^2(\lambda; D^{INP})$ let us denote as $\Lambda^2_{NP}$,
then a symbolic relation between them and $\Lambda^2_{QCD} \equiv
\Lambda^2_{PT}$ could be written as

\begin{equation}
\Lambda^2_{NP} \longleftarrow^{\infty \leftarrow \alpha_s}_{0
\leftarrow M_{IR}} \ \Delta^2_{JW} \ { }^{\alpha_s \rightarrow
0}_{M_{UV} \rightarrow \infty} \longrightarrow  \ \Lambda^2_{PT}.
\end{equation}
Here $\alpha_s$ is obviously the fine structure coupling constant
of strong interactions, while $M_{UV}$ and $M_{IR}$ are the UV and
IR cut-offs, respectively. The right-hand-side limit is well known
as the weak coupling regime, and we know how to take it within the
renormalization group equations approach \cite{1,2,6,7,8,9}.
However, it would be interesting to understand within our approach
how the JW mass gap $\Delta^2_{JW}$ may actually become
$\Lambda^2_{PT}$. The left-hand-side limit can be regarded as the
strong coupling regime, and we hope that we have explained here
how to begin to deal with it, not solving the gluon SD equation
directly, which is a formidable task, anyway. However, there is no
doubt that the final goal of this limit, namely, the mass gap
$\Lambda_{NP}$ exists, and should be renormalization group
invariant in the same way as $\Lambda_{QCD}$. It is solely
responsible for the large-scale structure of the true QCD ground
state, while $\Lambda_{PT}$ is responsible for the nontrivial PT
dynamics there.

Evidently, such kind of relation (7.1) is only possible due to the
explicit presence of a mass gap in the gluon SD equation of
motion. It leads to the exact separation between the truly NP and
nontrivial PT parts (phases) at the level of a single gluon
propagator. It also provides a basis for the restoration of the
transversality of the gluon propagator relevant for NP QCD by
formulating the above-described and demonstrated subtraction
prescription. A possible relation between these two phases shown
in Eq. (7.1) is a manifestation that "the problems encountered in
perturbation theory are not mere mathematical artifacts but rather
signify deep properties of the full theory" \cite{43}. The message
that we are trying to convey is that the nontrivial PT phase in
the full gluon propagator indicates the existence of the truly NP
one (the INP one within the general iteration solution) and the
other way around.

A few years ago Jaffe and Witten have formulated the following
theorem \cite{5}:

\vspace{3mm}

 {\bf Yang-Mills Existence And Mass Gap:} Prove that
for any compact simple gauge group $G$, quantum Yang-Mills theory
on $\bf{R}^4$ exists and has a mass gap $\Delta > 0$.

\vspace{3mm}

Of course, at present to prove the existence of the YM theory with
compact simple gauge group $G$ is a formidable task yet. It is
rather mathematical than physical problem. However, in the case of
acceptance of our proposal, one of the main results here can be
then formulated similar to the above-mentioned JW theorem as
follows:

\vspace{3mm}

{\bf Mass Gap Existence:} If quantum Yang-Mills theory with
compact simple gauge group $G=SU(3)$ exists on $\bf{R}^4$, then it
has a mass gap $\Delta > 0$.

\vspace{3mm}

It is important to emphasize that a mass gap has not been
introduced by hand. It is hidden in the skeleton loop integrals,
contributing to the gluon self-energy, and dynamically generated
mainly due to the NL interaction of massless gluon modes. No
truncations/approximations and no special gauge choice are made
for the above-mentioned regularized skeleton loop integrals. An
appropriate subtraction scheme has been applied to make the
existence of a mass gap perfectly clear. Within the general
iteration solution the mass gap shows up explicitly when the gluon
momentum goes to zero. The Lagrangian of QCD does not contain a
mass gap, while it explicitly appears in the gluon SD equation of
motion. This once more underlines the importance of the
investigation of the SD system of equations and identities
\cite{1,2,8,9} for understanding the true structure of the QCD
ground state. We have established the structure of the regularized
full gluon propagator (see Eqs. (3.16) and (3.17)) and the
corresponding SD equation (3.22) in the presence of a mass gap.

In order to realize a mass gap, we propose not to impose the
transversality condition on the gluon self-energy (see Eq. (3.6),
while preserving the color gauge invariance condition (3.15) for
the full gluon propagator. This proposal is justified by the NL
and NP dynamics of QCD (the constant skeleton tadpole contribution
to the gluon self-energy explicitly violates its transversality
structure). Such a temporary violation of color gauge
invariance/symmetry (TVCGI/S) is completely NP effect, since in
the PT limit $\Delta^2=0$ this effect vanishes. Let us emphasize
that we would propose this even if there were no explicit
violation of the transversality of the gluon self-energy by the
constant skeleton tadpole term. In other words, whether this term
is explicitly present or not, but just color confinement (the
gluon is not a physical state) gives us a possibility not to
impose the transversality condition on the gluon self-energy. The
existence of this term is a hint that the above-mentioned
transversality might be temporary violated. Since the gluon is not
a physical state because of color confinement as mentioned above,
the TVCGI/S in QCD has no direct physical consequences. None of
physical quantites/processes in low-energy QCD will be directly
affected by this proposal.

For the calculations of physical observables from first principles
in low-energy QCD we need the full gluon propagator, which
transversality has been sacrificed in order to realize a mass gap
(despite their general role the ghosts cannot guarantee its
transversality in this case). However, we have already pointed out
how the transversality of the gluon propagator relevant for NP QCD
is to be restored at the final stage. In accordance with our
prescription it becomes automatically transversal, free of the PT
contributions ("contaminations"), and it regularly depends on the
mass gap, so that it vanishes when the mass gap goes to zero. The
role of the first necessary subtraction (6.4) at the fundamental
gluon propagator level or (5.19) in the case of INP QCD is to be
emphasized. We also briefly described some other types of the
subtractions at the hadronic level as well (i.e., when gluon and
quark degrees of freedom are to be integrated out).

In QED a mass gap is always in the "gauge prison". It cannot be
realized even temporarily, since the photon is a physical state.
However, in QCD a door of the "color gauge prison" can be opened
for a moment in order to realize a mass gap, because the gluon is
not a physical state. A key to this "door" is the constant
skeleton tadpole term. On the other hand, this "door" can be
opened without key (as any door) by not imposing the
transversality condition on the gluon self-energy. So in QED a
mass gap cannot be "liberated" from the vacuum, while photons and
electrons can be liberated from the vacuum in order to be physical
states. In QCD a mass gap can be "liberated" from the vacuum,
while gluons and quarks cannot be liberated from the vacuum in
order to be physical states. In other words, there is no breakdown
of $U(1)$ gauge symmetry in QED because the photon is a physical
state. At the same time, a temporary breakdown of $SU(3)$ color
gauge symmetry in QCD is possible because the gluon is not a
physical state (color confinement).

Let us emphasize one more that no truncations/approximations and
no special gauge have been made for the corresponding skeleton
loop integrals within our approach, i.e., it is pure NP, by its
nature. So on the general ground we have established the existence
at least of two different types of solutions for the full gluon
propagator in the presence of a mass gap. The so-called general
iteration solution (5.18) is always severely singular in the IR
($q^2 \rightarrow 0$), i.e., the gluons always remain massless,
and this does not depend on the gauge choice (this behavior of the
full gluon propagator in different approximations and gauges has
been earlier obtained and investigated in many papers, see, for
example Ref. \cite{10} and references therein). The massive-type
solution (4.6) leads to an effective gluon mass, which explicitly
depends on the gauge-fixing parameter, and it cannot be directly
identified with the mass gap. Moreover, we were unable to make an
effective gluon mass a gauge-invariant as a result of the
renormalization, and therefore to assign to it a physical meaning.
This solution becomes smooth at $q^2 \rightarrow 0$ in the Landau
gauge $\xi=0$ only. Both types of solutions are independent from
each other and should be considered on equal footing, since the
gluon SD equation is highly NL system. For such kind of systems
the number of solutions is not fixed $a \ priori$. The UV behavior
($q^2 \rightarrow \infty$) of all solutions should be fixed by AF
\cite{1}. Due to unsolved yet confinement problem, the IR behavior
($q^2 \rightarrow 0$) is not fixed. Only solution of the color
confinement problem will decide which type of formal solutions
really takes place. At the present state of arts none of them can
be excluded \cite{44}.

In summary, the behavior of QCD at large distances is governed by
a mass gap, possibly realized in accordance with our proposal. The
dynamically generated mass gap is usually related to breakdown of
some symmetry (for example, the dynamically generated quark mass
is an evidence of chiral symmetry breakdown). Here a mass gap is
an evidence of the TVCGI/S. In the presence of a mass gap the
coupling constant becomes play no role. This is also a direct
evidence of the "dimensional transmutation", $g^2 \rightarrow
\Delta^2(\lambda, \alpha, \xi, g^2)$ \cite{1,45,46}, which occurs
whenever a massless theory acquires masses dynamically. It is a
general feature of spontaneous symmetry breaking in field
theories. The mass gap has to play a crucial role in the
realization of the quantum-dynamical mechanism of color
confinement \cite{5}.

\begin{acknowledgments}

Support in part by HAS-JINR and Hungarian OTKA-T043455 grants (P.
Levai) is to be acknowledged. The author is grateful to A.
Kacharava, N. Nikolaev, P. Forgacs, L. Palla, J. Nyiri, and
especially to C. Hanhart and A. Kvinikhidze for useful discussions
and remarks during his stay at IKP (Juelich).

\end{acknowledgments}

\appendix
\section{Renormalization group equation for the effective
charge}

It is instructive to make some preliminary remarks, concerning
solution of the renormalization group equation for the regularized
effective charge, which appears in the general iteration solution
(5.18). This equation leads to the determination of the
corresponding $\beta$-function, and it is

\begin{equation}
q^2 {d \alpha_s(q^2) \over dq^2} = \beta(\alpha_s(q^2)).
\end{equation}
As we have already established, the effective charge for this
solution can be uniquely decomposed as the exact sum of the two
principally different terms, namely

\begin{equation}
\alpha_s(q^2) = \alpha_s(q^2)- \alpha^{PT}_s(q^2)+
\alpha^{PT}_s(q^2) = \alpha^{INP}_s(q^2) + \alpha^{PT}_s(q^2),
\end{equation}
where the explicit expression for the INP part of the effective
charge is given by Eq. (6.13), which is valid in the whole
energy/momentum range. We also omit the dependence on the mass gap
in the effective charge as unimportant. Let us remind that the PT
part of the effective charge is a regular function at small $q^2$,
so it can be explicitly present as the corresponding Taylor
expansion

\begin{equation}
\alpha^{PT}_s(q^2) = \sum_{k=0}^{\infty} ( q^2/\mu^2)^k
\alpha_s^{(k)}(0),
\end{equation}
where $\mu^2$ is some fixed mass squared parameter introduced in
Eq. (5.12). In principle the coefficients of the Taylor expansion
$\alpha_s^{(k)}(0)$ depend on the same set of parameters as the
coefficients $\Phi_k(\lambda, \alpha, \xi, g^2)$ of the Laurent
expansion (6.13), i.e., $\alpha_s^{(k)}(0) = \alpha_s^{(k)} \equiv
\alpha_s^{(k)}(\lambda, \alpha, \xi, g^2)$. For convenience, let
us introduce short-hand notation $\Phi_k \equiv \Phi_k(\lambda,
\alpha, \xi, g^2)$ as well. Then on account of Eq. (6.13), the
effective charge (A2) explicitly becomes

\begin{equation}
\alpha_s(q^2) = \alpha^{INP}_s(q^2) + \alpha^{PT}_s(q^2) =
{\Delta^2 \over q^2} \sum_{k=0}^{\infty} (\Delta^2 / q^2)^k \Phi_k
+ \sum_{k=0}^{\infty} (q^2/\mu^2)^k \alpha_s^{(k)}.
\end{equation}
Substituting this expression  into the renormalization group
equation (A1), one obtains the formal solution for the
corresponding $\beta$-function as follows:

\begin{equation}
\beta(\alpha_s(q^2)) = - {\Delta^2 \over q^2}
\sum_{k=0}^{\infty}(1+k) (\Delta^2 / q^2)^k  \Phi_k +
\sum_{k=0}^{\infty} ( q^2/\mu^2)^k k \alpha_s^{(k)},
\end{equation}
so the $\beta$-function can be also exactly and uniquely
decomposed into the two different terms, namely

\begin{equation}
\beta(\alpha_s(q^2)) = \beta^{INP}(\alpha^{INP}_s(q^2)) +
\beta^{PT}(\alpha^{PT}_s(q^2)),
\end{equation}
where

\begin{equation}
\beta^{INP}(\alpha^{INP}_s(q^2)) = - {\Delta^2 \over q^2}
\sum_{k=0}^{\infty}(1+k) (\Delta^2 / q^2)^k \Phi_k  = -
\alpha^{INP}_s(q^2) - {\Delta^2 \over q^2} \sum_{k=0}^{\infty}
(\Delta^2 / q^2)^k k \Phi_k,
\end{equation}
and

\begin{equation}
\beta^{PT}(\alpha^{PT}_s(q^2)) = \sum_{k=0}^{\infty} (q^2/\mu^2)^k
k \alpha_s^{(k)}.
\end{equation}
Let us note that in the NP region (i.e., at small $q^2$) from Eq.
(A4) one obtains

\begin{equation}
\alpha_s(q^2) = \alpha^{INP}_s(q^2) + O(1), \quad q^2 \rightarrow
0,
\end{equation}
while from Eqs. (A6) and (A8) it follows

\begin{equation}
\beta(\alpha_s(q^2)) = \beta^{INP}(\alpha^{INP}_s(q^2)) + O(q^2),
\quad q^2 \rightarrow 0.
\end{equation}
Thus, one can conclude that in the NP region the $\beta$-function
of the general iteration solution as a function of its argument is
determined by its INP part, which is always in the domain of
attraction (i.e., negative, see Eq. (A7)) as it is required for
the confining theory \cite{1}.

\end{document}